\documentstyle[12pt,epsfig]{article}
\newskip\humongous \humongous=0pt plus 1000pt minus 1000pt

\newif\ifdtup

\def\pr#1{#1^\prime}

\def\beq{\begin{equation}}
\def\eeq{\end{equation}}

\def\beqn{\begin{eqnarray}}
\def\eeqn{\end{eqnarray}}
\relax

\def\dotx{\dotx{\dot\overline{x}}}

\relax

\catcode`\@=11

\@addtoreset{equation}{section}
\def\theequation{\thesection\arabic{equation}}

\def\@normalsize{\@setsize\normalsize{15pt}\xiipt\@xiipt
\abovedisplayskip 14pt plus3pt minus3pt%
\belowdisplayskip \abovedisplayskip
\abovedisplayshortskip \z@ plus3pt%
\belowdisplayshortskip 7pt plus3.5pt minus0pt}

\def\small{\@setsize\small{13.6pt}\xipt\@xipt
\abovedisplayskip 13pt plus3pt minus3pt%
\belowdisplayskip \abovedisplayskip
\abovedisplayshortskip \z@ plus3pt%
\belowdisplayshortskip 7pt plus3.5pt minus0pt
\def\@listi{\parsep 4.5pt plus 2pt minus 1pt
     \itemsep \parsep
     \topsep 9pt plus 3pt minus 3pt}}

\@twosidetrue

\relax

\catcode`@=12

\catcode`\@=11

\def\section{\@startsection{section}{1}{\z@}{3.5ex plus 1ex minus
   .2ex}{2.3ex plus .2ex}{\large\bf}}

\def\thesection{\arabic{section}.}

\def\appendix{\setcounter{section}{0}
 \def\thesection{APPENDIX \Alph{section}:}
 \def\theequation{\Alph{section}.\arabic{equation}}}

\def\ps@headings{\def\@oddfoot{}\def\@evenfoot{}
\def\@oddhead{\hbox{}\hfill
 \makebox[.5\textwidth]{\raggedright\ignorespaces --\thepage{}--
 \hfill {}}}  
\def\@evenhead{\@oddhead}
\def\subsectionmark##1{\markboth{##1}{}}
}

\ps@headings

\catcode`\@=12

\def\figcap{\section*{Figure Captions\markboth
 {FIGURECAPTIONS}{FIGURECAPTIONS}}\list
 {Fig. \arabic{enumi}:\hfill}{\settowidth\labelwidth{Fig. 999:}
 \leftmargin\labelwidth
 \advance\leftmargin\labelsep\usecounter{enumi}}}
 \relax
\def\tablecap{\section*{Table Captions\markboth
 {TABLECAPTIONS}{TABLECAPTIONS}}\list
 {Table \arabic{enumi}:\hfill}{\settowidth\labelwidth{Table 999:}
 \leftmargin\labelwidth
 \advance\leftmargin\labelsep\usecounter{enumi}}}
 \relax
\def\reflist{\section*{References\markboth
 {REFLIST}{REFLIST}}\list
 {[\arabic{enumi}]\hfill}{\settowidth\labelwidth{[999]}
 \leftmargin\labelwidth
 \advance\leftmargin\labelsep\usecounter{enumi}}}
 \relax

\catcode`\@=11

\def\ps@headings{\def\@oddfoot{}\def\@evenfoot{}
\def\@oddhead{\hbox{}\hfill
 \makebox[.5\textwidth]{\raggedright\ignorespaces --\thepage{}--
 \hfill {}}}    
\def\@evenhead{\@oddhead}
\def\subsectionmark##1{\markboth{##1}{}}
}

\ps@headings

\relax

\relax
\def\pl#1#2#3{{\it Phys. Lett. }{\bf #1}(19#2)#3}

\def\prl#1#2#3{{\it Phys. Rev. Lett. }{\bf #1}(19#2)#3}

\def\prep#1#2#3{{\it Phys. Rep. }{\bf #1}(19#2)#3}
\def\pr#1#2#3{{\it Phys. Rev. }{\bf #1}(19#2)#3}

\relax

\catcode `@ 11
\def\biblabel#1{\if@filesw\immediate
\write\@auxout{\string\bibcite{#1}{\the\value{\@listctr }}}\fi}
\catcode `@ 12

\newcommand{\ccaption}[2]{
  \begin{center}
    \parbox{0.85\textwidth}{
      \caption[#1]{\small\it {#2}}}
  \end{center}    }
\def    \be             {\begin{equation}}
\def    \ee             {\end{equation}}
\def    \ba             {\begin{eqnarray}}
\def    \ea             {\end{eqnarray}}

\def    \=              {\;=\;}
\def    \frac           #1#2{{#1 \over #2}}

\def \as   {\ifmmode \alpha_s \else $\alpha_s$ \fi}

\def\b0{b_0}

\def \mt   {\ifmmode m_{\rm t} \else $m_{\rm t}$ \fi}

\def \mur  {\mbox{$\mu_{\rm \scriptscriptstyle{R}}$}}
\def \muf  {\mbox{$\mu_{\rm \scriptscriptstyle{F}}$}}

\begin{document}
\begin{titlepage}
\nopagebreak
{\flushright{
        \begin{minipage}{4cm}
        CERN-TH/96-21  \hfill \\
        hep-ph/9602208\hfill \\
        \end{minipage}        }

}
\vfill
\begin{center}
{\LARGE { \bf \sc  The Top Cross Section \\[0.3cm]
           in Hadronic Collisions}}
\vskip .5cm
{\bf Stefano CATANI\footnote{Research supported in part by the EEC programme
``Human Capital and Mobility'', Network ``Physics at High Energy Colliders'',
contract CHRX-CT93-0357(DG 12 COMA).}}
\\
\vskip .1cm
{INFN, Sezione di Firenze and Univ. di Firenze, Florence, Italy} \\
\vskip .5cm
{\bf Michelangelo L. MANGANO\footnote{On leave of absence
from INFN, Pisa, Italy},}
{\bf Paolo NASON\footnote{On leave of absence from INFN, Milan, Italy},}
\\
\vskip 0.1cm
{CERN, TH Division, Geneva, Switzerland} \\
\vskip .5cm
{\bf Luca TRENTADUE$^1$}
\\
\vskip .1cm
{Univ. di Parma and INFN, Gruppo Collegato di Parma, Parma, Italy}
\end{center}
\nopagebreak
\vfill
\begin{abstract}
  We reexamine the top quark production cross section at the Tevatron and LHC,
  in the light of recent progress on the resummation of
  logarithmic soft gluon corrections.
  We find that resummation effects are much smaller than previously
  thought. We also compute Coulombic threshold effects, and find them
  negligible.
  We update the discussion of uncertainties due to scale dependence,
  the value of the
  strong coupling constant, and the parton density parametrization.
  Our current best estimate of the top production cross section at the Tevatron
  and its error is $\sigma(\mt=175\,{\rm GeV})=4.75{+0.73\atop-0.62}$.
\end{abstract}
\vskip 1cm
CERN-TH/96-21  \hfill \\
\today \hfill
\vfill
\end{titlepage}
\section{Introduction}
We present in this letter a theoretical reassessment of the evaluation of the
top quark production cross section in high-energy hadronic collisions. The
top quark having been found \cite{cdf,d0},
the comparison between its observed
production properties and those expected from the Standard Model will be an
important probe of the possible existence of new phenomena.
One of the most important tests to be performed concerns the total production
cross section. This is the most inclusive quantity available, and is a
priori the least sensitive to a detailed understanding of the higher order
corrections which influence the evolution of the initial and final states.
Within QCD, one expects the perturbative expansion in powers of
$\as(\mt)$ to be well behaved and to provide an accurate estimate of the
total cross section already at low orders.
In particular, the first estimates of the total production cross section using
the full next-to-leading-order (NLO) matrix elements~\cite{sigtot,gual,kellis}
gave an increase (relative to the Born result) of the order
of 30\% for masses above 100 GeV. The residual perturbative QCD uncertainty,
evaluated by
varying the renormalization and factorization scales, was shown to be no larger
than 10\%. The choice of parametrization for the input parton densities was
also shown to give effects of this order of magnitude,
by using the available sets.

It was
pointed out in ref.~\cite{sigres} that logarithmic contributions
associated to the
emission of soft gluons from the initial state could significantly enhance the
NLO result. 
An independent study of the soft gluon resummation in top production has
appeared recently~\cite{BergerContopanagos}.
These studies are based on the
resummation formulae obtained in the pioneering works of
refs.~\cite{largexresum,Sterman,CataniTrentadue}. Using these
formulae for explicit calculations
requires the choice of a prescription in order to
bypass the Landau-pole singularity which is exposed by the integration
of the QCD running coupling over
gluon energies of the order of $\Lambda_{QCD}$.
In particular it was shown in ref.~\cite{sigres} that,
within the proposed formalism,
one can get sensible results only by cutting off the soft gluon
emission at scales below $0.1\div 0.2\; \mt$.
The choice made in ref.~\cite{BergerContopanagos} follows instead
a suggestion put forward in
ref.~\cite{StermanContopanagos}. According to this prescription the Landau pole
appearing in the Mellin transform of the resummed hard cross section should be
integrated over in a principal value sense.
It was argued in
ref.~\cite{StermanContopanagos} that this prescription results in
non-perturbative power-suppressed ambiguities scaling like $\Lambda_{QCD}/Q$ in
the hard scale $Q$ typical of the process under consideration.
The net effect of gluon resummation evaluated in
ref.~\cite{BergerContopanagos}
amounts to an increase by approximately 10\% of the NLO cross section.

In a companion paper \cite{cmnt2} we will show that the
approaches of refs.~\cite{sigres,BergerContopanagos}
overestimate both the gluon resummation contribution and
the associated non-perturbative residual uncertainty.
This is a consequence of the fact that their resummation formulae introduce
unjustified factorially growing terms in the perturbative expansion.
In ref. \cite{cmnt2} we will also show that a more natural prescription
exists, in which these terms are not present.
In the following, we will briefly summarize the basis
of our criticism and the definition and main properties of the new proposed
resummation prescription. We will then present a numerical study indicating
that the  impact of resummation on the total
top production cross section is of the
order of a percent, much smaller than previously thought. We will also
show that threshold corrections due to higher order Coulomb
effects are likewise negligible.
We will conclude this study with an updated analysis of the current theoretical
uncertainties coming from the scale dependence, from the
choice of partonic densities and from the value of \as.

It should be pointed out that this study is performed within the strict domain
of the Standard Model. Corrections to the top cross section much larger than
the Standard Model QCD uncertainties can be
obtained in the presence of new phenomena. For a
partial list of specific examples, see {\it e.g.}
refs.~\cite{lane-et-al}.

\section{The minimal prescription for the resummation}
The heavy flavour production cross section in the Born
approximation is given by the formula
\beq \label{HVQCrossSection}
\sigma=\sum_{i,j=q\bar{q},\bar{q}q,gg}
\int_0^1 dx_1\,dx_2 F^{(1,i)}(x_1) F^{(2,j)}(x_2)\;\hat{\sigma}^{(ij)}
\left(\frac{\rho}{x_1\,x_2}\right)\;,\quad\rho=\frac{4m^2}{S}\;,
\eeq
where $m$ is the mass of the heavy quark, and $S$ is the square of the
total centre-of-mass energy.
The functions $F^{(i,k)}$ are the parton densities for parton $k$
in hadron $i$, and are evaluated at a scale $\mu$ of the order of the
heavy quark mass $m$.
Explicit formulae for the partonic cross section
$\hat\sigma$ can be found for example in ref.~\cite{sigtot}.
The corresponding $N$ space formula is
\beq \label{MPHQ}
\sigma=\sum_{i,j=q\bar{q},\bar{q}q,gg} \frac{1}{2\pi i}
\int_{C-i\infty}^{C+i\infty}\; F^{(1,i)}_{N+1}\;F^{(2,j)}_{N+1}\;
\hat\sigma^{(ij)}_N\; \rho^{-N}\; dN\;,
\eeq
where the Mellin transforms are defined in the following way
\beq
F_N=\int_0^1 \frac{dx}{x} x^N F(x)\,,
\quad \hat\sigma_N=\int_0^1 \frac{dz}{z} z^N\hat\sigma(z)\,.
\eeq
The integration contour lies to the right of all singularities.

We use the following resummed formula
\beq\label{sigmares}
\sigma^{(\rm res)}=\sum_{i,j=q\bar{q},\bar{q}q,gg} \frac{1}{2\pi i}
\int_{C_{\rm MP}-i\infty}^{C_{\rm MP}+i\infty}\;
F^{(1,i)}_{N+1}\;F^{(2,j)}_{N+1}\;
\Delta^{(ij)}_{N+1} \hat\sigma^{(ij)}_N\; \rho^{-N}\; dN\;.
\eeq
where, in the $\overline{\rm MS}$ scheme
\beq\label{deltahf}
\ln \Delta^{(ij)}_N = \ln N \;g_{ij, \,1}(\b0\as\ln N)
 + {\cal O}(\as^k \ln^{k}N)
\eeq
where
\beq\label{g1hf}
g_{q{\bar q}, \,1}(\lambda) = g_1^{\overline {\rm MS}}(\lambda)
\;\;, \;\;\;\;
g_{gg, \,1}(\lambda) = \frac{C_A}{C_F} \;g_1^{\overline {\rm MS}}(\lambda)
\eeq
\beq\label{g0ms}
g_1^{{\overline {\rm MS}}}(\lambda) =
+\frac{C_{\rm F}}{\pi\b0\lambda}\Bigl[ \,2\lambda + (1-2\lambda)
\ln(1-2\lambda) \Bigr] \;,
\eeq
and
\beq
b_0=\frac{11C_A-2n_f}{12\pi}\,\quad C_A=3\,,\quad C_F=\frac{4}{3}\;.
\eeq
In formula (\ref{deltahf}) the strong coupling constant $\as$ is
evaluated at a scale $\mu$ of the order of the heavy quark mass $m$.
The factor $\ln\Delta_N^{(ij)}$ resums all the leading logarithmic
terms $\as^k\,\ln^{k+1}N$ due to soft gluon emission.
The contour in eq.~(\ref{sigmares}) is chosen
between the cuts on the negative $N$ axis and the singularity
at $b_0\as\log N=1/2$, which is a Landau pole.
We will call this the minimal prescription (MP). In ref.~\cite{cmnt2} we
will show that this prescription enjoys the following remarkable properties:
\begin{itemize}
\item It is accurate at the leading log level in the threshold limit.
\item If we expand the MP formula in powers of $\alpha(m)$, the
expansion is asymptotic to the full formula.
\item The power expansion of the MP formula does not have any factorially
growing coefficients, and therefore is free of ambiguities
of the order of powers of $\Lambda/m$.
The ambiguity associated with the asymptotic
expansion of the MP resummation is exponentially suppressed, being of the order
$e^{-B(1-\rho)\frac{m}{\Lambda}}$.
\end{itemize}
We claim that previous calculations of resummation effects
\cite{sigres,BergerContopanagos}
have a corresponding power expansion which does have
factorially growing terms. These terms are spurious. They arise either
from certain approximations performed
when going from the $N$ space to the $x$ space formulae, or from
the integration of the running coupling constant down to the
Landau singularity.

Our result is consistent with the findings of ref.~\cite{BB}, indicating that
no power-like $1/Q$ ambiguity emerges from the resummation of multiple gluon
emission in Drell--Yan processes.

A full discussion of these issues will appear in a companion paper
\cite{cmnt2}, where we will also study the implications 
of this prescription for
Drell--Yan, heavy quark and jet production in hadronic collisions.

In the present work we will report on results obtained with our
MP formula of eq.~(\ref{sigmares}), which was implemented without any further
approximation in a numerical program.

\section{Results}
In this section we present the results obtained for a very heavy
quark, and in particular for the
top cross section.
First of all, we remind the reader that large soft gluon effects
are present not only in the production cross section, but also
in the deep-inelastic processes that are used to determine the structure
functions. In order to perform a complete study of the resummation
one should therefore use resummed formulae also when fitting deep-inelastic
scattering data, direct photon production data, and in general all
phenomena which are used to constrain the parton densities.
In the present work we will assume that the structure functions
have been properly extracted from data (including the resummation effects)
and assess the significance of the resummation for heavy flavour production.
Effects of comparable size may arise from refitting the structure
functions using resummed formulae. Our conclusions will mostly be based
upon the fact that the resummation effects we find are in fact very
small, and can be neglected.

\begin{figure}
\centerline{\epsfig{figure=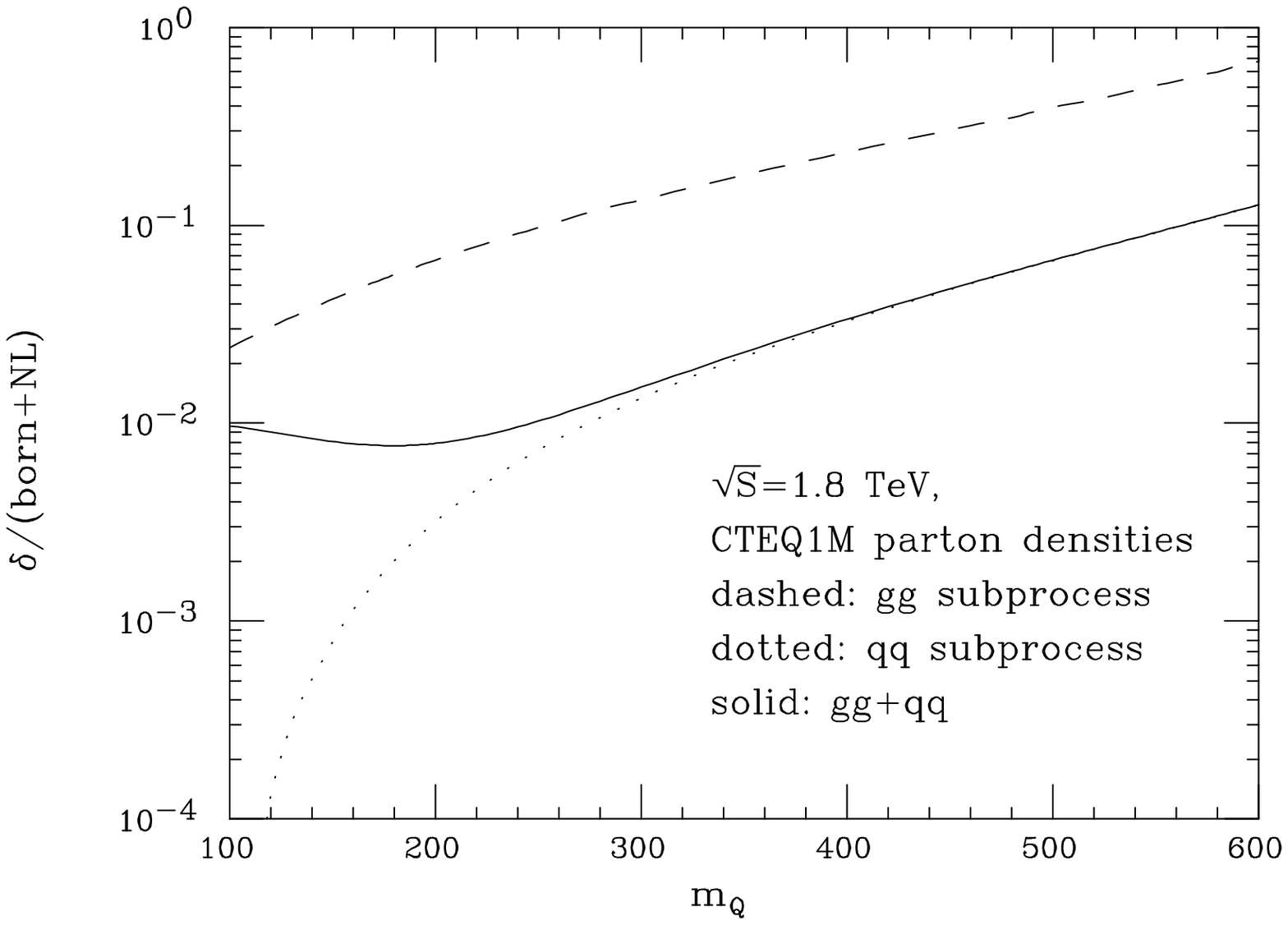,width=0.7\textwidth,clip=}}
\ccaption{}{ \label{frtev}
Contribution of gluon resummation at order $\as^4$ and higher, relative to the
NLO result, for the individual subprocesses and for the total,
as a function of the top mass in $p\bar p$ collisions at 1.8 TeV. }
\end{figure}

The importance of the resummation
effects is illustrated in fig.~\ref{frtev}
where we plot the quantities
\be
\frac{\delta_{\rm gg}}{\sigma^{(gg)}_{\rm NLO}}\,,\quad
\frac{\delta_{\rm q\bar{q}}}{\sigma^{(q\bar{q})}_{\rm NLO}}\,,\quad
\frac{\delta_{\rm gg}+\delta_{\rm q\bar{q}}}{\sigma^{(gg)}_{\rm NLO}+
\sigma^{(q\bar{q})}_{\rm NLO}}
\ee
where $\delta$ is equal to
our MP resummed hadronic cross section~(\ref{sigmares})
in which the terms of order $\as^2$
and $\as^3$ have been subtracted,
and $\sigma_{(\rm NLO)}$ is the full hadronic NLO cross section~\cite{sigtot}.
In other words, what we display is the relative contribution to the total cross
section coming from corrections of order higher than those accounted for by the
full NLO result. We show the results for the $q\bar q$ and $gg$ production
channels separately, as well as for the total. The results are shown
at $\sqrt{S}=1.8\;$TeV as a function of the top mass
in the range $100 < \mt({\rm GeV}) < 600$.
The wide mass
range is chosen for the sole purpose of illustration. We chose a common
renormalization and factorization scale $\mu=\mt$, and parton densities from
the CTEQ1 set \cite{cteq1}.

The plot clearly indicates
that the contribution of soft-gluon resummation is negligible, unless the top
mass approaches the total hadronic energy. This result is consistent with
the expectation that the soft
gluon enhancement of the production cross sections should be relevant
only very close to threshold.
For a top mass of $175$ GeV the increase due to this effect is below 1\%.
We also note that the
effect of resummation is potentially much larger for the $gg$ channel,
because of the $C_{\rm A}/C_{\rm F}$ enhancement in the exponent 
of the resummation formula (see eq. \ref{g1hf}).
It turns out, however, that when the $gg$ subprocess becomes important
({\it i.e.} for small top masses) we are so far from threshold that the
resummation effect is very small. Conversely, for large masses,
the quark component, which has smaller resummation effects,
dominates.

\begin{figure}
\centerline{\epsfig{figure=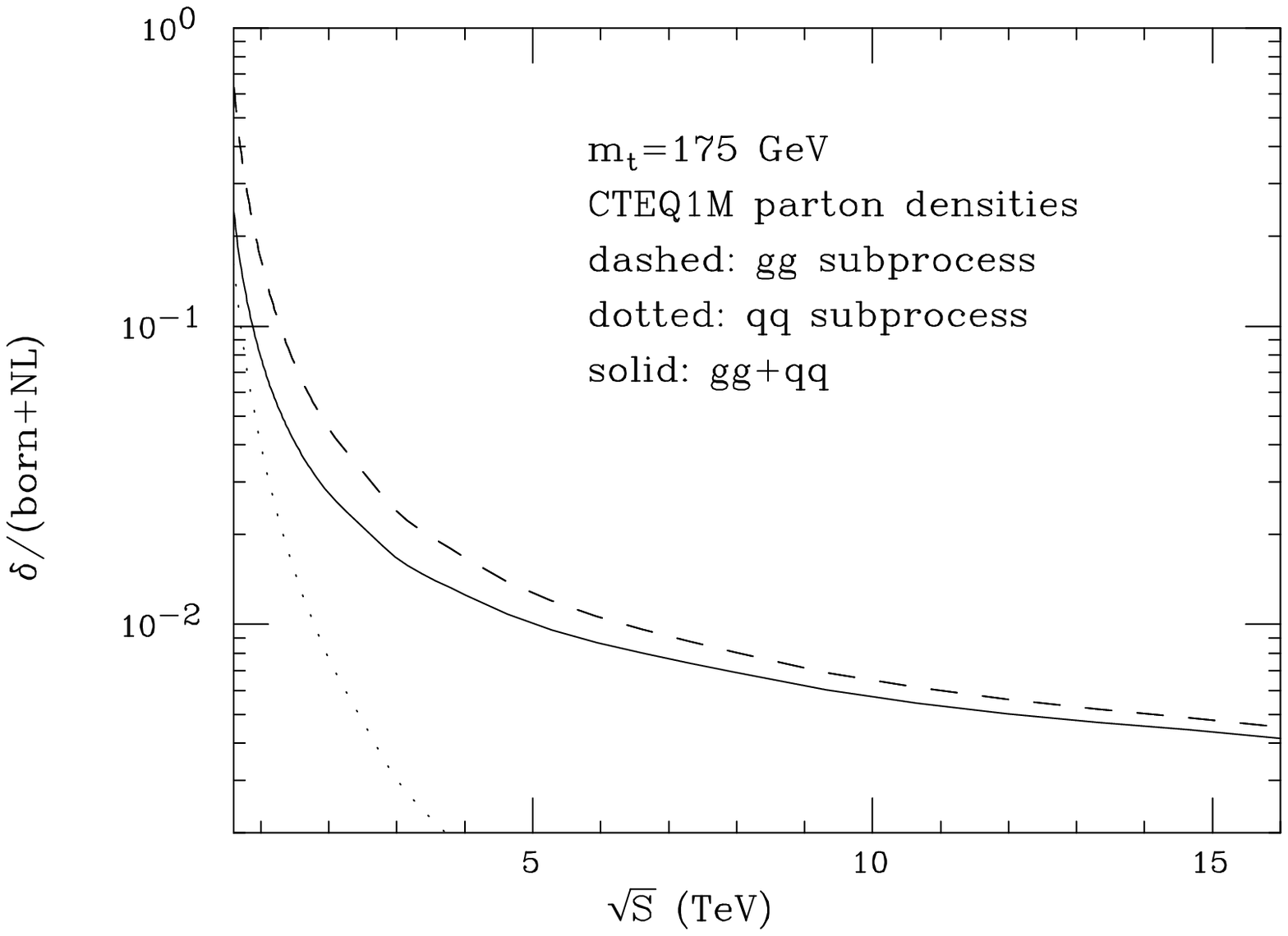,width=0.7\textwidth,clip=}}
\ccaption{}{ \label{frlhc}
Contribution of gluon resummation at order $\as^4$ and higher, relative to the
NLO result, for the individual channels and for the total,
as a function of the CM energy in $pp$ collisions. }
\end{figure}

In fig.~\ref{frlhc} we show the same quantities, at fixed top mass,
versus the CM energy.
We see that at the LHC,
where production is dominated by the $gg$ initial
state, we are far enough from the hadronic threshold, so that the
resummation effects are again small.
We conclude therefore that the effects of gluon resummation can be
neglected at the present stage, as they are much smaller than the uncertainties
present in the NLO determination.
Note that, due to the dominance of the $gg$ subprocess, the relative
importance of the resummation
effects at $\sqrt{S}=1.8$ TeV is larger in $pp$ collisions
than in $p\bar{p}$.

\begin{figure}
\centerline{\epsfig{figure=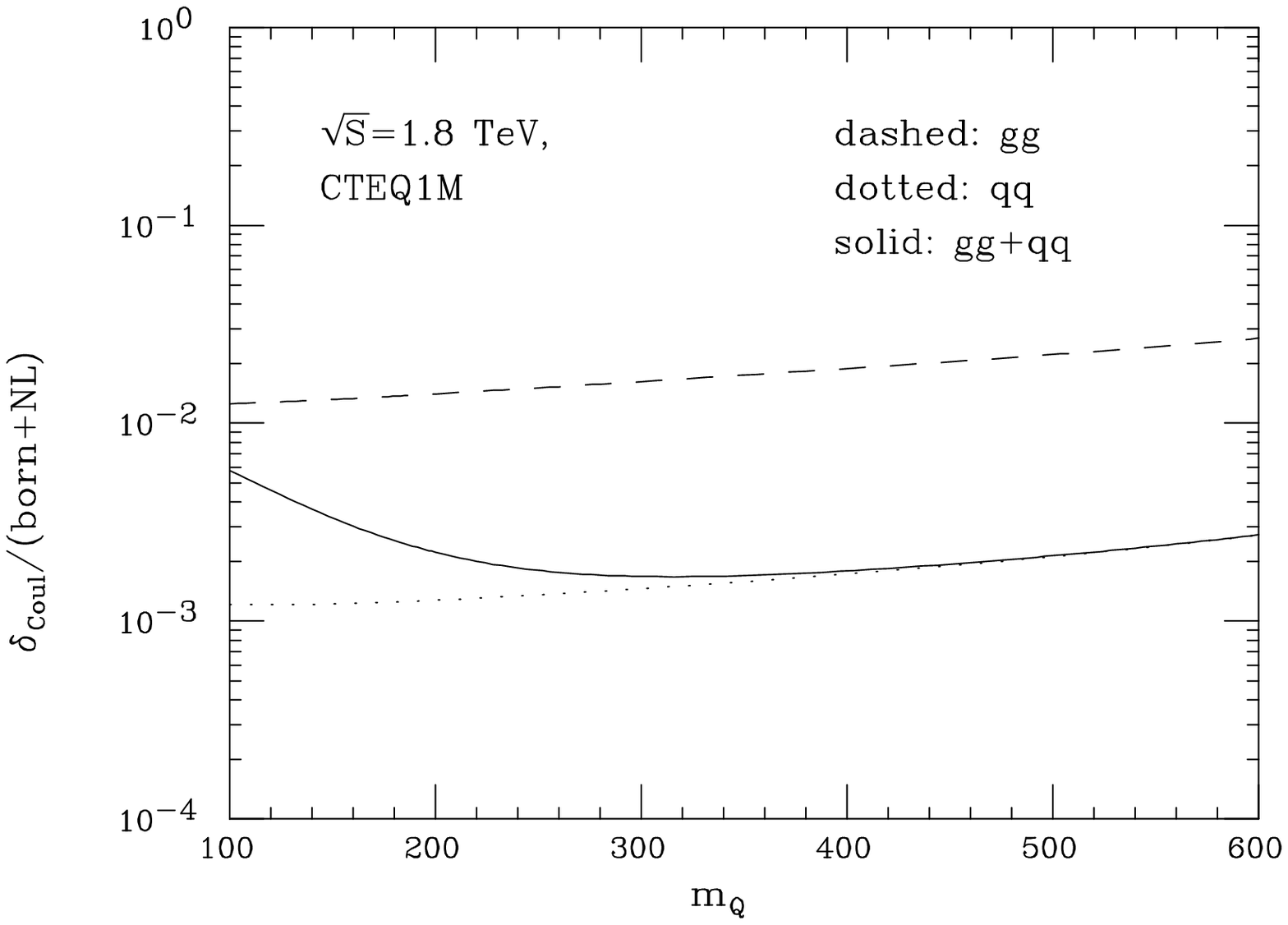,width=0.7\textwidth,clip=}}
\ccaption{}{ \label{topcoul}
Contribution of Coulomb effects at order $\as^4$ and higher, relative to the
NLO result, for the individual channels and for the total,
as a function of the CM energy in $p\bar{p}$ collisions. }
\end{figure}

Beside the soft gluon emission effects, also Coulomb effects
may enhance or deplete the cross section near threshold
\cite{Sommerfeld,Fadin}.
We have calculated these effects in the following way.
We have separated the partonic Born cross section formulae into their colour
singlet and octet components
\beqn
\hat{\sigma}^{(gg)}&=&\hat{\sigma}_{(8)}^{(gg)}+\hat{\sigma}_{(1)}^{(gg)}
\\
\hat{\sigma}^{(q\bar{q})}&=&\hat{\sigma}_{(8)}^{(q\bar{q})}
\eeqn
where $\hat{\sigma}^{(q\bar{q},gg)}$ can be found in refs.~\cite{sigtot}, and
\beq
\hat{\sigma}_{(1)}^{(gg)}(\rho)=\frac{\as^2}{m^2}\;
\frac{\beta\rho\pi}{384}\left[\frac{1}{\beta}\log\frac{1+\beta}{1-\beta}
(4+4\rho-2\rho^2)-4-4\rho\right]\;,
\eeq
where $\beta=\sqrt{1-\rho}$.
The Coulomb-resummed cross section is given as
\beq
\hat\sigma^{\rm Coul}(\rho)
=\hat\sigma_{(8)}(\rho)\frac{\pi\as/(6\beta)}{\exp(\pi\as/(6\beta))-1}
+\hat\sigma_{(1)}(\rho)
\frac{4\pi\as/(3\beta)}{1-\exp(-4\pi\as/(3\beta))}\;.
\eeq
We do not include bound state effects, which, as shown in ref.~\cite{Fadin},
are much smaller.
In fig.~\ref{topcoul} we have plotted the hadronic quantity
\beq
\frac{\delta_{\rm Coul}}{\sigma_{(\rm NLO)}}
\eeq
which (similarly to the case of soft gluon effects)
represents the relative correction due to the resummation
of Coulomb effects not already included into the NLO results.
We see that these effects are similar in magnitude to the soft-gluon
effects, and fully negligible at the Tevatron.
Since at the LHC we are further away from the threshold region,
we conclude that also there they will be negligible.
Therefore, the following estimates of the top cross
section will not include neither soft gluon effects,
nor Coulombic ones. Electroweak corrections to top production in
hadronic collisions have been considered in refs.~\cite{EWtop}.
They range from -0.97\% to -1.74\% of the born cross section,
for a higgs mass of 60 and 1000 GeV
respectively ~\cite{EWtop1}. They will also not be included in the
following estimates.

\begin{figure}
\centerline{\epsfig{figure=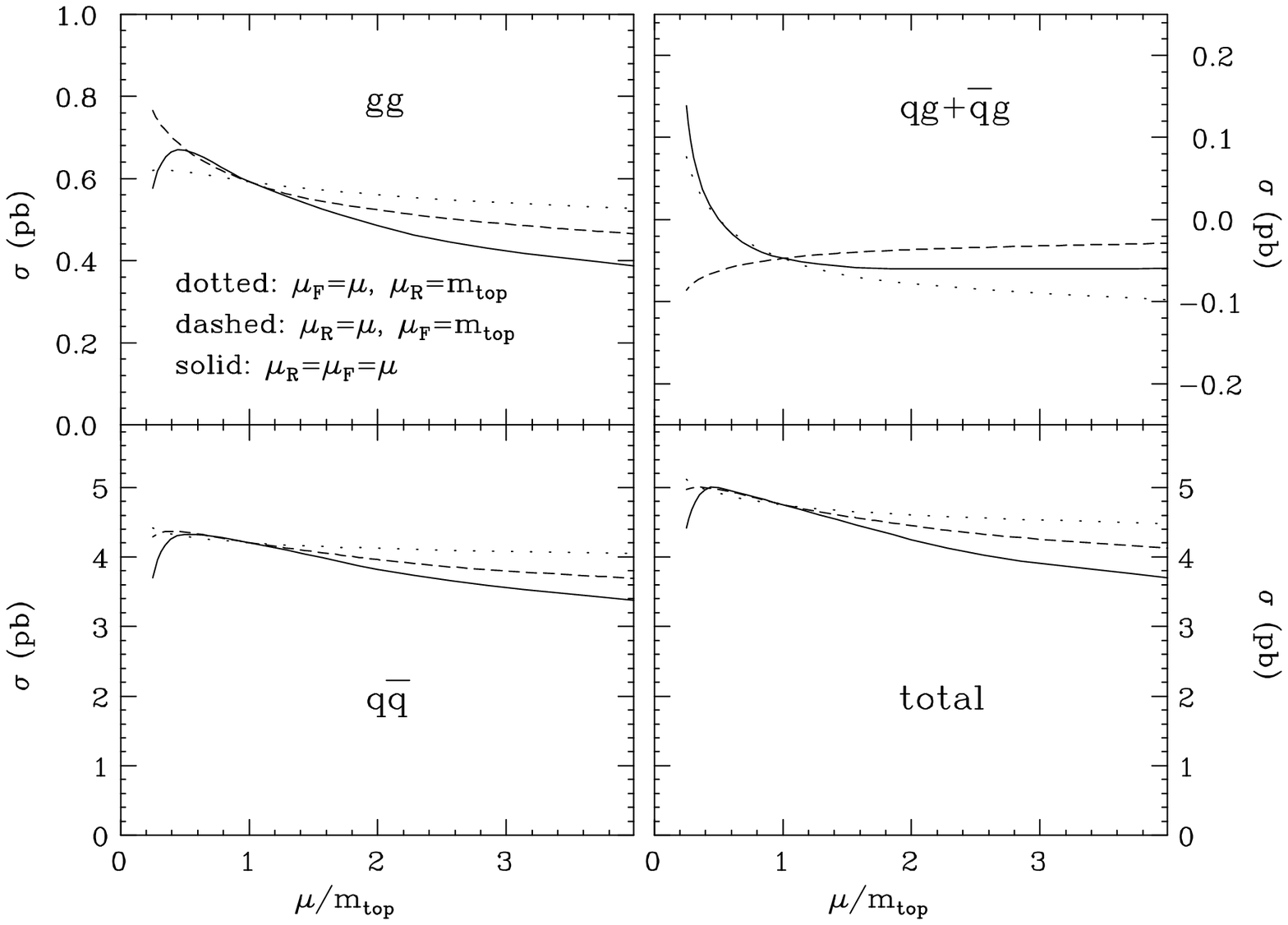,width=0.7\textwidth,clip=}}
\ccaption{}{ \label{scales}
Scale dependence of the top cross section at the Tevatron.
Dotted lines are the $\mu_{\rm F}$ dependence at fixed  $\mu_{\rm R}$,
dashed lines are the $\mu_{\rm R}$ dependence at fixed  $\mu_{\rm F}$,
and solid lines are obtained by the simultaneous variation of $\mu_{\rm R}$
and $\mu_{\rm F}$.
}
\end{figure}

We now proceed to review the current theoretical
uncertainties that arise at NLO. Uncertainties due to
unknown higher-order effects are usually accounted for by varying
the renormalization ($\mu_{\rm R}$) and factorization
($\mu_{\rm F}$) scales. In principle, independent
variations of the two scales should be considered. 
In fig.~\ref{scales} we show the scale dependence of the top
cross section.
Note that there is compensation of the factorization scale dependence
when the different subprocesses are added up.
In particular, the $gg$ and $qg$ factorization
scale dependence have opposite behaviour. In fact, it is only the combined
scale dependence that can be considered an estimate of the neglected
subleading corrections. As a second point, we observe that
the maximum of the cross section is reached around $m_t/2$. Thus, the usual
choice of the range $m_t/2<\mu<2m_t$
appears to be particularly justified in this case. As a third point,
we observe that the scale dependence in the cross section goes in the same
direction for the two scales, so that, for the purpose of estimating
the associated uncertainty, it is sufficient to consider the simultaneous
variation of $\mu_{\rm R}$ and $\mu_{\rm F}$.

\begin{figure}
\centerline{\epsfig{figure=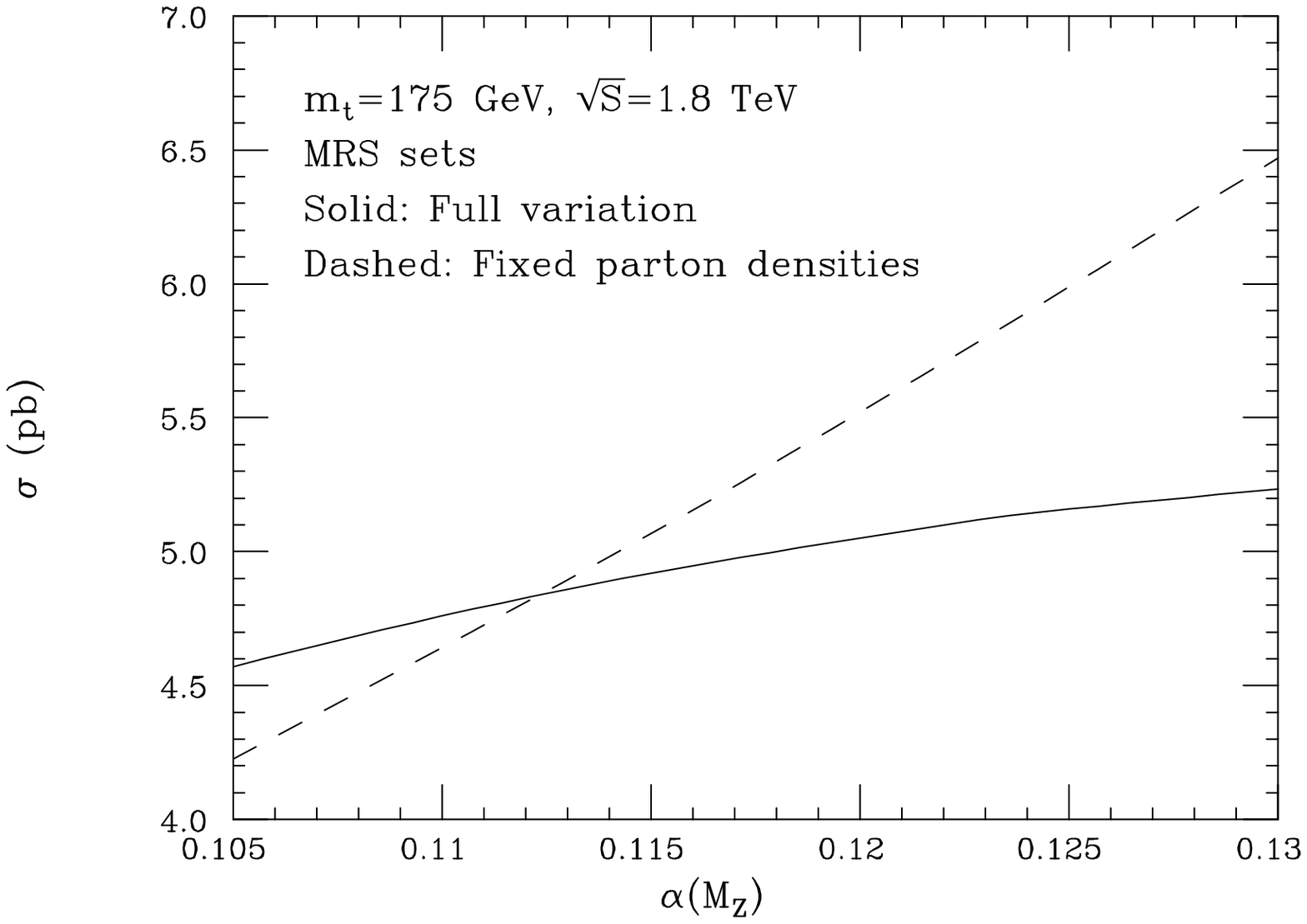,width=0.7\textwidth,clip=}}
\ccaption{}{ \label{lambdavar}
Top cross section as a function of $\as(M_{\rm Z})$.
The dotted line does not include the variation of the parton
densities due to the change in $\as$. }
\end{figure}

Aside from the scale uncertainties, which reflect the limitations of
the perturbative QCD calculation,
there are uncertainties associated to our imprecise
knowledge of the physical parameters involved. In particular,
the strong coupling constant is determined within a certain accuracy.
In the case of top production at the Tevatron, larger strong
couplings tend to give larger partonic cross section. However, for larger
strong couplings, the Altarelli--Parisi evolution,
which occurs along the wide span in $Q^2$ from the values at which
deep-inelastic fits are performed ($\approx 10\; {\rm GeV}^2$) to the top
mass, softens the quark parton densities, thereby decreasing the cross
section. In order to perform a fair estimate of the uncertainty due to
$\Lambda_{\rm QCD}$, we need sets of parton densities
fitted with different values
of the strong coupling. The sets of ref.~\cite{MRSLambda} meet our
purpose\footnote{The values of $\Lambda_4$ that accompany the fortran
program for the structure function sets of ref.~\cite{MRSLambda}
are not consistent with the values of $\as$ quoted there,
the differences being of the order of 1\%. In the present work,
we extract the values of $\Lambda_5$ from their quoted values of $\as$ using
the standard two-loop formula~\protect{\cite{PDG}}.}.
In fig.~\ref{lambdavar} we show the cross section as a function of the
strong coupling.
For comparison we also show the $\as$ dependence
if the parton densities are kept fixed. We see the remarkable reduction
in the  $\as$ dependence due to the compensation of the rise
of the partonic cross section and the decrease of the quark parton densities.

Our results for the top cross section are collected in tables \ref{top175},
\ref{top160} and \ref{top190}, for values of the top mass that span the current
uncertainties. We also show results obtained with with the recent
parametrization derived in ref.~\cite{cteqp} by including the CDF jet data.
We stress that the numbers in the table are obtained with a standard
two-loop calculation. No resummation effects are included.
{\renewcommand{\arraystretch}{1.8}
\begin{table}
\begin{center}
\begin{tabular}{|c|c|c|c|c|c|c|c|c|} \hline
& CTEQ1M & CTEQ$^\prime$ & MRSA$^\prime$ & \multicolumn{5}{c|}{MRS,
variable $\Lambda$,  $\as(M_{\rm Z})=$} \\ \cline{5-9}
 \mur=\muf   &   &       &       &$0.105$ &
$0.110$ & $0.115$ & $0.120$ &
$0.125$  
\\ \hline\hline
\mt/2 & 5.24 & 5.07 & 5.00 & 4.78 & 4.99 & 5.18 & 5.34 & 5.48
\\ \hline
\mt   & 4.96 & 4.86 & 4.75 & 4.57 & 4.76 & 4.92 & 5.05 & 5.16
\\ \hline
2\mt  & 4.38 & 4.38 & 4.25 & 4.13 & 4.27 & 4.38 & 4.47 & 4.52
\\ \hline
\end{tabular}
\ccaption{}{\label{top175} Total cross sections for \mt=175 GeV at NLO.}
\end{center}
\end{table} }

{\renewcommand{\arraystretch}{1.8}
\begin{table}
\begin{center}
\begin{tabular}{|c|c|c|c|c|c|c|c|c|} \hline
& CTEQ1M & CTEQ$^\prime$ & MRSA$^\prime$ &
\multicolumn{5}{c|}{MRS, variable $\Lambda$, $\as(M_{\rm Z})=$}
\\ \cline{5-9}
  \mur=\muf  &    &       &       &$0.105$ &
$0.110$ & $0.115$ & $0.120$ &
$0.125$  
\\ \hline\hline
\mt/2 & 8.64 & 8.18 & 8.19 & 7.80 & 8.17 & 8.50 & 8.81 & 9.06
\\ \hline
\mt   & 8.16 & 7.85 & 7.78 & 7.45 & 7.78 & 8.06 & 8.32 & 8.53
\\ \hline
2\mt  & 7.21 & 7.08 & 6.95 & 6.72 & 6.97 & 7.18 & 7.35 & 7.48
\\ \hline
\end{tabular}
\ccaption{}{\label{top160} Total cross sections for \mt=160 GeV at NLO.}
\end{center}
\end{table} }

{\renewcommand{\arraystretch}{1.8}
\begin{table}
\begin{center}
\begin{tabular}{|c|c|c|c|c|c|c|c|c|} \hline\hline
& CTEQ1M & CTEQ$^\prime$ & MRSA$^\prime$ &
\multicolumn{5}{c|}{MRS, variable $\Lambda$, $\as(M_{\rm Z})=$} \\ \cline{5-9}
 \mur=\muf &      &       &       &$0.105$ &
$0.110$ & $0.115$ & $0.120$ &
$0.125$  
\\ \hline\hline
\mt/2 & 3.26 & 3.22 & 3.13 & 3.00 & 3.13 & 3.24 & 3.32 & 3.39
\\ \hline
\mt   & 3.08 & 3.08 & 2.97 & 2.87 & 2.98 & 3.07 & 3.14 & 3.20
\\ \hline
2\mt  & 2.72 & 2.77 & 2.66 & 2.59 & 2.67 & 2.73 & 2.78 & 2.78
\\ \hline
\end{tabular}
\ccaption{}{\label{top190} Total cross sections for \mt=190 GeV at NLO.}
\end{center}
\end{table} }
Our ranges for the top cross section are thus given by
$7.78{+1.28\atop-1.06}$~pb, $4.75{+0.73\atop-0.62}$~pb,
$2.97{+0.42\atop-0.38}$~pb
for \mt=160, 175, 190 GeV respectively,
and are adequately fitted in this range by the expression
\beq
\sigma(t\bar{t})=e^{\frac{175-m_{\rm t}}{31.5}}
(\mbox{$4.75{+0.73\atop-0.62}$})\,.
\eeq
As a central value for our determination we have chosen the MRSA$^\prime$
\cite{MRSAp} result with $\mu_{\rm R}=\mu_{\rm F}=m_{\rm t}$, in association
with a value of $\Lambda_5=0.152\,$GeV (which corresponds to
$\as(M_Z)=0.1113$, according to the standard two-loop formula \cite{PDG}).
For the MRS sets with variable $\Lambda$, we have used
$\Lambda_5=0.0994,\,.140,\,.190,\,.253,\,.328\,$GeV (which corresponds to
$\as(M_Z)=0.105,\,0.110,\,0.115,\,0.120,\,0.125$).

\begin{figure}
\centerline{\epsfig{figure=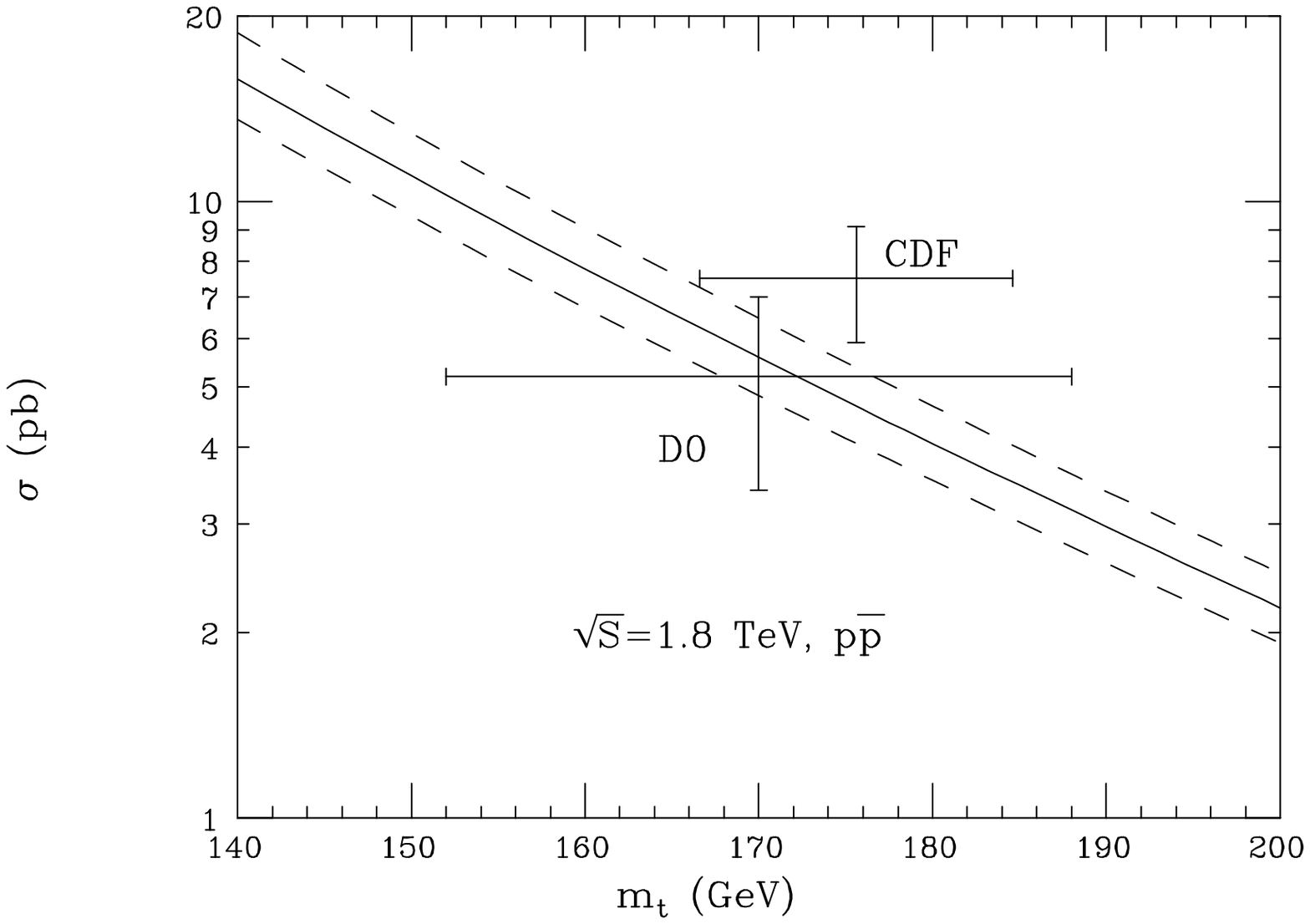,width=0.7\textwidth,clip=}}
\ccaption{}{ \label{bandtop1}
Top cross section at the Tevatron at $\sqrt{S}=1.8\;$TeV.
The solid line is obtained with MRSA$^\prime$
parton density, and the dashed lines correspond to the upper and
lower values obtained in tables~1--3. The experimental
data are taken from ref.~\protect{\cite{expxsect}}.}
\end{figure}

\begin{figure}
\centerline{\epsfig{figure=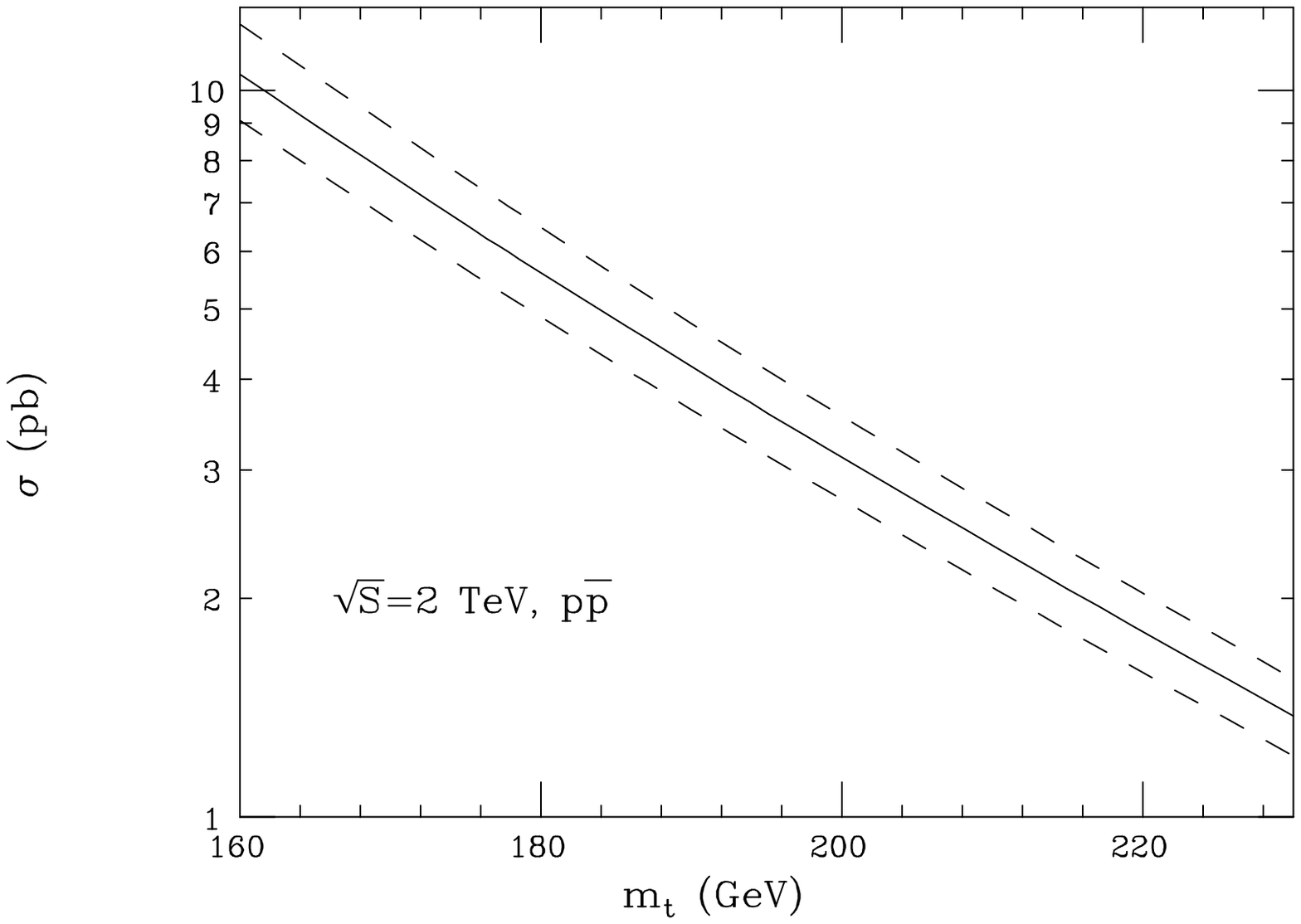,width=0.7\textwidth,clip=}}
\ccaption{}{ \label{bandtop2}
Top cross section at the Tevatron at $\sqrt{S}=2\;$TeV. }
\end{figure}

In $p\bar{p}$ collisions at $\sqrt{S}=2\;$TeV we obtain, with the same method,
a top cross section of $10.5{+1.8\atop-1.4}$~pb, $6.53{+1.03\atop-0.86}$~pb,
$4.17{+0.61\atop-0.53}$~pb for \mt=160, 175, 190 GeV respectively.
The cross section bands are also shown in fig.~\ref{bandtop1},
for $\sqrt{S}=1.8\;$TeV and fig.~\ref{bandtop2} for $\sqrt{S}=2\;$TeV. 
Because of the high precision of this theoretical prediction,
the measurement of the top cross section could become a sensitive probe
of new physics \cite{lane-et-al} \cite{topsusy}.

For reference, we also quote the cross section for top production at the
LHC. We get $\sigma(t\bar{t})=0.77{+0.25\atop -0.12}\;$nb,
for $m_t=175\;$GeV
and $\sqrt{S}=14\;$TeV. The error is obtained with the same scale and $\as$
variations we used for the Tevatron
cross section. 
\vskip 0.5 cm
\noindent
{\bf Acknowledgements}

We wish to thank R.K. Ellis for useful comments.
We also wish to thank W. Hollik and D. Wackeroth for computing
the value of the electroweak corrections to top production quoted above.
Two of us (SC and LT)  acknowledge the hospitality and the partial support
received by the CERN TH division while this work was carried out.
                                                  

\clearpage
\begin{thebibliography}{99}
\def    \nuke   #1#2#3{{\sl Nucl. Phys.} {\bf B#1}  (#2) #3}
\def    \pl     #1#2#3{{\sl Phys. Lett.} {\bf #1B}  (#2) #3}
\def    \prl    #1#2#3{{\sl Phys. Rev. Lett.} {\bf #1}  (#2) #3}
\def    \pr     #1#2#3{{\sl Phys. Rev.} {\bf #1}  (#2) #3}
\def    \prd    #1#2#3{{\sl Phys. Rev.} {\bf D#1}  (#2) #3}
\def    \prep   #1#2#3{{\sl Phys. Rep.} {\bf #1}  (#2) #3}
\def    \zeit   #1#2#3{{\sl Z. Phys.} {\bf #1}  (#2) #3}
\bibitem{cdf}
  F. Abe et al., CDF Collab., \prl{74}{1995}{2626}; \\
  F. Abe et al., CDF Collab., \prd{50}{1994}{2966}.
\bibitem{d0}
  S. Abachi et al., D0 Collab., \prl{74}{1995}{2632}.
\bibitem{sigtot}
  P.~Nason, S.~Dawson and R.~K.~Ellis,
  \nuke{303}{1988}{607};
  W.~Beenakker, H. Kuijf, W.L. van Neerven and J. Smith,
  \prd{40}{1989}{54}.
\bibitem{gual}
  G. Altarelli, M. Diemoz, G. Martinelli and P. Nason,
  \nuke{308}{1988}{724}.
\bibitem{kellis}
  R.K. Ellis, \pl{259}{1991}{492}.
\bibitem{sigres}
  E. Laenen, J. Smith and W.L. van Neerven,
  \nuke{369}{1992}{543}; \pl{321}{1994}{254}.
\bibitem{BergerContopanagos}
  E. Berger and H. Contopanagos, \pl{361}{1995}{115};\\
  preprint ANL-HEP-CP-95-85, hep-ph/9512212.
\bibitem{largexresum}
  Yu.L. Dokshitzer, D.I. Dyakonov and S.I. Troyan, \prep{58}{1980}{271};\\
  G. Parisi, \pl{90}{1980}{295};\\
  G. Curci and M. Greco, \pl{92}{1980}{175};\\
  D. Amati et al., \nuke{173}{1980}{429};\\
  M. Ciafaloni and G. Curci, \pl{102}{1981}{352};\\
  P. Chiappetta, T. Grandou, M. Le Bellac and J.L. Meunier,
  \nuke{207}{1982}{251}.
\bibitem{Sterman}
  G. Sterman, \nuke{281}{1987}{310}.
\bibitem{CataniTrentadue}
  S. Catani and L. Trentadue, \nuke{327}{1989}{323};\newline
  \nuke{353}{1991}{183}.
\bibitem{StermanContopanagos}
  G. Sterman and H. Contopanagos, \nuke{419}{1994}{77};\\
  L. Alvero and H. Contopanagos, \nuke{436}{1995}{184}.
\bibitem{cmnt2}
  S. Catani, M.L. Mangano, P. Nason and L. Trentadue, in preparation.
\bibitem{lane-et-al}
  E. Eichten and K. Lane, \pl{327}{1994}{129};  \\
  C.T. Hill and S.J. Parke, \prd{49}{1994}{4454};\\
  R. Casalbuoni et al., CPT-95/P.3176, UGVA-DPT 1995/04-887,
  hep-ph/9505212;\\
  B. Holdom, M.V. Ramana, \pl{353}{1995}{295}.
\bibitem{BB}
  M. Beneke and V.M.~Braun, \nuke{454}{1995}{253};\\
  Yu.L. Dokshitser, G. Marchesini and B.R. Webber,
  preprint CERN-TH-95-281,  hep-ph/9512336. 
\bibitem{cteq1}
  J. Botts et al., \pl{304}{1993}{159}.
\bibitem{Sommerfeld}
  A. Sommerfeld ``Atombau und Spektrallinien'', Bd. 2 (Vieweg, Braunschweig,
  1939); A.D. Sakharov, JETP {\bf 18} (1948) 631.
\bibitem{Fadin}
  V. Fadin, V. Khoze and T. Sj\"ostrand, \zeit{C48}{1990}{613}.
\bibitem{EWtop}
  S. Willenbrock and A. Stange, \prd{48}{1993}{2054};\\
  W. Beenakker, A. Denner, W. Hollik, R. Mertig, T. Sack and D. Wackeroth,
   \nuke{411}{1994}{343};\\
  C. Kao, G.A. Ladinsky, C.P. Yuan FSU-HEP-930508, hep-ph/9305270;\\
  R. Harlander, M. Jezabeck and J.H. K\"uhn, TTP 95-25, hep-ph/9506292.
\bibitem{EWtop1}
  W. Hollik and D. Wackeroth, private communication.
\bibitem{MRSLambda}
  A.D. Martin, R.G. Roberts and W.J. Stirling,
  \pl{356}{1995}{89}.
\bibitem{PDG}
  L. Montanet et al., \prd{50}{1994}{1173}
  and 1995 off-year partial update for the 1996 edition available on 
  the PDG WWW pages (URL: http://pdg.lbl.gov/).
\bibitem{cteqp}
  J. Huston et al., Michigan State Preprint MSU-HEP-50812, hep-ph/9511386.
\bibitem{MRSAp}
  A.D. Martin, R.G. Roberts and W.J. Stirling, \prd{50}{1994}{6734}.
\bibitem{topsusy}
  J.M. Yang and C.S. Li, \prd{52}{1995}{1541};\\
  C.S. Li, B.Q. Hu, J.M. Yang and C.G. Hu, \prd{52}{1995}{1541}.
\bibitem{expxsect} 
  A. Caner, for the CDF Collaboration, presented at the
  10$^{\rm th}$ Les Rencontres de Physique de la Vall\'ee d'Aoste, La Thuile,
  Val d'Aosta, March 3-9, 1996;\newline
  M. Narain, for the D0 Collabration, presented at the
  10$^{\rm th}$ Les Rencontres de Physique de la Vall\'ee d'Aoste, La Thuile,
  Val d'Aosta, March 3-9, 1996.
\end{thebibliography}
\end{document}